\documentclass[twocolumn,prl,showpacs,floatfix,superscriptaddress]{revtex4}

\usepackage{amsmath,amssymb,amsfonts,bm}
\usepackage[usenames,dvipsnames]{color}
\usepackage{graphicx}
\usepackage{multirow}
\usepackage{tabularx}

\newcommand{\Deltam}{\bm{\Delta}}
\newcommand{\Hm}{\mathbf{H}}
\newcommand{\cm}{\mathbf{c}}

\begin{document}

\title{Phases of attractive spin-imbalanced fermions in square lattices}

\author{Simone Chiesa}
\affiliation{Department of Physics, College of William \& Mary,
Williamsburg, VA 23188, USA} 

\author{Shiwei Zhang}
\affiliation{Department of Physics, College of William \& Mary,
Williamsburg, VA 23188, USA} 

\begin{abstract}

We determine the relative stability of different ground-state phases 
of spin-imbalanced populations of attractive fermions in square lattices. 
The ground state is determined within Hartree-Fock-Bogoliubov theory, with care taken to remove finite size effects.
The phases are systematically characterized by the symmetry of the order parameter and 
their real- and momentum-space structures.
For quarter- to half-filled lattices, where the Fermi surfaces are most distorted
from their spherical counterpart in the continuum, we predominantly find
unidirectional Larkin-Ovchinikov-type phases.
We discuss the effect of commensuration between
the ordering wave vector and the density imbalance,
and describe the mechanism of Fermi surface reconstruction and pairing
for various orders. A robust supersolid phase exists 
when the ordering wave vector is diagonally directed. Charge and
pairing order coexist, rather than competing, and are responsible
for the opening of the gap on different portions of the Fermi surface.
A variational determination of the correct pair
momentum of the Larkin-Ovchinikov phases shows that phase separation
does not occur in the considered regime of density and magnetization.
\end{abstract}

\maketitle

\section{Introduction}
There has been a surge of interest in the possibility of realizing
unconventional fermionic superfluids using cold atomic gases. Amongst 
the many possibilities offered by the highly tunable Hamiltonians
available in cold-atoms experiments, the simplest remains
that of unequal populations of two hyperfine states in the presence
of attractive interaction.
The theoretical study of such systems dates back to Fulde and Ferrel (FF) \cite{Fulde1964} and 
Larkin and Ovchinikov (LO) \cite{Larkin1965} who independently suggested that the mismatch between 
the Fermi surfaces of the two species could result in the formation of a condensate 
of finite-momentum pairs. 
Atomic gases offer a direct route to the realization of FFLO phases, circumventing most of the
difficulties of solid state systems, thanks to the possibility of 
controlling independently the density of the two species, the absence of disorder
and, most importantly, the ability to engineer strong interactions.
In spite of this, the existence of an FFLO phase in three dimensions
has been argued to be confined to a small 
range of interaction strengths and polarizations\cite{Sheehy2006},
and detection has remained elusive.

The possibility of using optical lattices has been suggested by several
authors\cite{Moreo2007,Wolak2012} as a key ingredient to observe FFLO-type states.
The best empirical indication that this may be the case is 
provided by experiments on strongly-correlated-electrons materials and the fact 
that, when doped, these systems show a tendency toward formation of inhomogeneities
in the form of spin, charge and, possibly, pairing density waves.
The relevance of these experiments to the properties of attractive fermions in optical lattices
stems from the belief that, in both cases, the essential physics can be captured
by a one-band Hubbard model: with an on-site repulsive interaction for many of the electronic systems
and an on-site attraction in an optical lattice. The attractive and repulsive
cases are mapped into each other by a particle-hole transformation\cite{Moreo2007} and the presence of 
spin-texture in the doped repulsive case translates into the occurrence of a modulated 
superfluid in an imbalanced population of attractive fermions, {\em i.e.} an FFLO phase.
This is reinforced by recent quantum Monte Carlo results \cite{Chang2010} on the 
two-dimensional repulsive model showing spin-density waves with 
long wavelength modulation. 

Despite this mapping and several works addressing the 
existence of a possible FFLO phase \cite{Koponen2006,Koponen2007,Loh2009}, 
the nature of the ground state phases in a spin-imbalanced two-dimensional optical lattice 
remains largely undetermined.
On the one hand, information on the repulsive model is entirely 
confined to the case of unpolarized systems, which maps
into the attractive case at half-filling: $n_\uparrow+n_\downarrow=1$; for the case of imbalanced
fermionic population, one is interested in the more general case of a 
polarized system and arbitrary
density.
On the other hand, works addressing the physics in the lattice
have either focused on the single plane-wave form of
the order parameter\cite{Koponen2006,Koponen2007} or on selected 
states\cite{Loh2009,Wang2006} (e.g. in
fixed size supercells) because of the challenge of removing large finite-size effects. Such restrictions can bias the result and lead to, for example, an incorrect pair momentum.
The accurate determination of the spatial structure
of the order parameter is also indispensable to addressing
the issue of phase separation.

In this work we establish the correct form that FFLO phases have in the thermodynamic limit
on a square lattice, and show that a proper determination of the leading pairing wave vectors in the 
ordered state leads to a characterization of different physical regimes
based on the properties of the nodal (excess) particles. 
Small to moderate interaction strengths are considered, where mean-field
theory is expected to capture the correct physics. In regime of density and
polarization where the presence of a lattice alters most dramatically the
shape of the Fermi surfaces, we find unidirectional order and the existence of three
distinct phases: 1) a nodal metallic state characterized by the presence
of Fermi arcs, 2) a nodal band insulator where the densities of excess particles and
nodal lines are equal and 3) a charge density wave that results in a robust
supersolid phase obtained when the ordering wave vector is directed along
the diagonal direction. 
Once the variational search for the optimal pairing wave vector includes
the proper symmetry of the order parameter, the Hartree-Fock-Bogoliubov ground state 
does not phase separate in the parameter regime of interest.

\section{Hartree-Fock-bogoliubov theory}
Results presented in this work are obtained using Hartree-Fock-Bogoliubov theory so 
that modulations in charge, spin and pairing are all handled on the 
same footing. The starting Hamiltonian reads
\begin{equation}
H= -t \sum_{\langle ij\rangle\sigma} c_{i\sigma}^\dagger c_{j\sigma} - U
\sum_i \big(n_{i\uparrow} n_{i\downarrow} - \mu n_i - \frac{h}{2} m_i\big),
\label{HubMod}
\end{equation}
where $c_{i\sigma}$ are fermionic annihilation operators of spin $\sigma$ on site $i$,
$n_{i\sigma}=c^\dagger_{i\sigma} c_{i\sigma}$, $n_i=n_{i\uparrow}+n_{i\downarrow}$ and
$m_i=n_{i\uparrow}-n_{i\downarrow}$.

In order to accommodate the inhomogeneities, the calculations are performed on 
supercells whose shape is dictated by the symmetry of the targeted phase. The supercell is
characterized by two basis vectors, $L_1$
and $L_2$, whose components are integers. Once the supercell shape is chosen, 
we define Bloch states as 
$c_j(k) \propto \sum_{L} c_{j+L} \exp\big[ik\cdot L\big]$
where
$L=n_1 L_1 + n_2 L_2$,
and $k$ is a vector that varies freely within the super-lattice first Brillouin zone.
Then, using these states and the mean-field approximation, the Hamiltonian 
decouples into a sum of $k$-dependent pieces, $H = \sum_k H(k)$, of the form
\begin{equation}
\begin{split}
H(k)= [\cm^\dagger_\uparrow \cm_\downarrow] 
\left[ \begin{array}{cc}
\Hm_\uparrow(k) & \Deltam  \\
\Deltam^\dagger & -\Hm_\downarrow^T(G-k)  \end{array} \right]
[\cm_\uparrow \cm^\dagger_\downarrow]^T\,,
\end{split}
\label{BdGH}
\end{equation}
where $\mathbf{c}_\uparrow$ and $\mathbf{c}_\downarrow$
represent an array (row) of operators, $\{c_{i\uparrow}(k)\}$ and $\{c_{i\downarrow}(G-k)\}$
respectively, with index $i$
running over the $N$ sites of the supercell, and $G$ is defined below.
$\Hm$ and $\Deltam$ are $N\times N$
matrices with elements
\begin{equation}
\begin{split}
[\Hm_\sigma(k)]_{ij} & = -t_{ij}(k) + \delta_{ij} ( D_{i\sigma} - \mu -s_\sigma h/2) \\
[\Deltam]_{ij}    & = \delta_{ij} \Delta_i .
\end{split}
\end{equation}
In the above, $t_{ij}(k)=\sum_L \exp(ik\cdot L) t_{i,j+L}$, $s_{\uparrow/\downarrow} = \pm 1$
and $D_{i\sigma}$, $\Delta_i$, $\mu$ and $h$ are determined by the
requirement that the Free energy 
$F = \langle H\rangle  - T S$
is a minimum for the target average densities $n_\sigma$.
This amounts to imposing the 
following self-consistency conditions
\begin{equation}
\begin{split}
D_{i\, -\sigma} =&\ -U \int dk \langle c^\dagger_{i\sigma}(k)c_{i\sigma}(k) \rangle\,, \\
\Delta_i        =&\ -U \int dk \langle c_{i\downarrow}(k)c_{i\uparrow}(k) \rangle\,,\\
n_\sigma  =&\  N^{-1} \sum_i \int dk \langle c^\dagger_{i\sigma}(k)c_{i\sigma}(k)\rangle\,,
\end{split}
\label{gapEq}
\end{equation}
with expectation values evaluated in the supercell. Note that, although we 
target specific densities, the mean-field approach works in the
grand-canonical ensemble. 

The variable $G$ in Eq.~(\ref{BdGH}) is a vector such that $\theta=G\cdot L$ gives the twist angle 
of the pairing order parameter under translation by $L$. For collinear phases, $\theta$ is either $0$
or $\pi$,
giving periodic or anti-periodic boundary conditions on $\langle c^\dagger_{i\uparrow} c^\dagger_{i\downarrow} \rangle$.
Note that uniform phases with a spiral order parameter
(FF type) amount to the choice 
of a one-site cell and a $G$ equal to the wave-vector of the spiral modulation.
Thus $H(k)$ is specified by a $2\times 2$ matrix, and the 
eigenvalues and eigenvectors can be computed analytically
in the spiral phase\cite{Fulde1964}.

\begin{figure}
\includegraphics[width=\columnwidth]{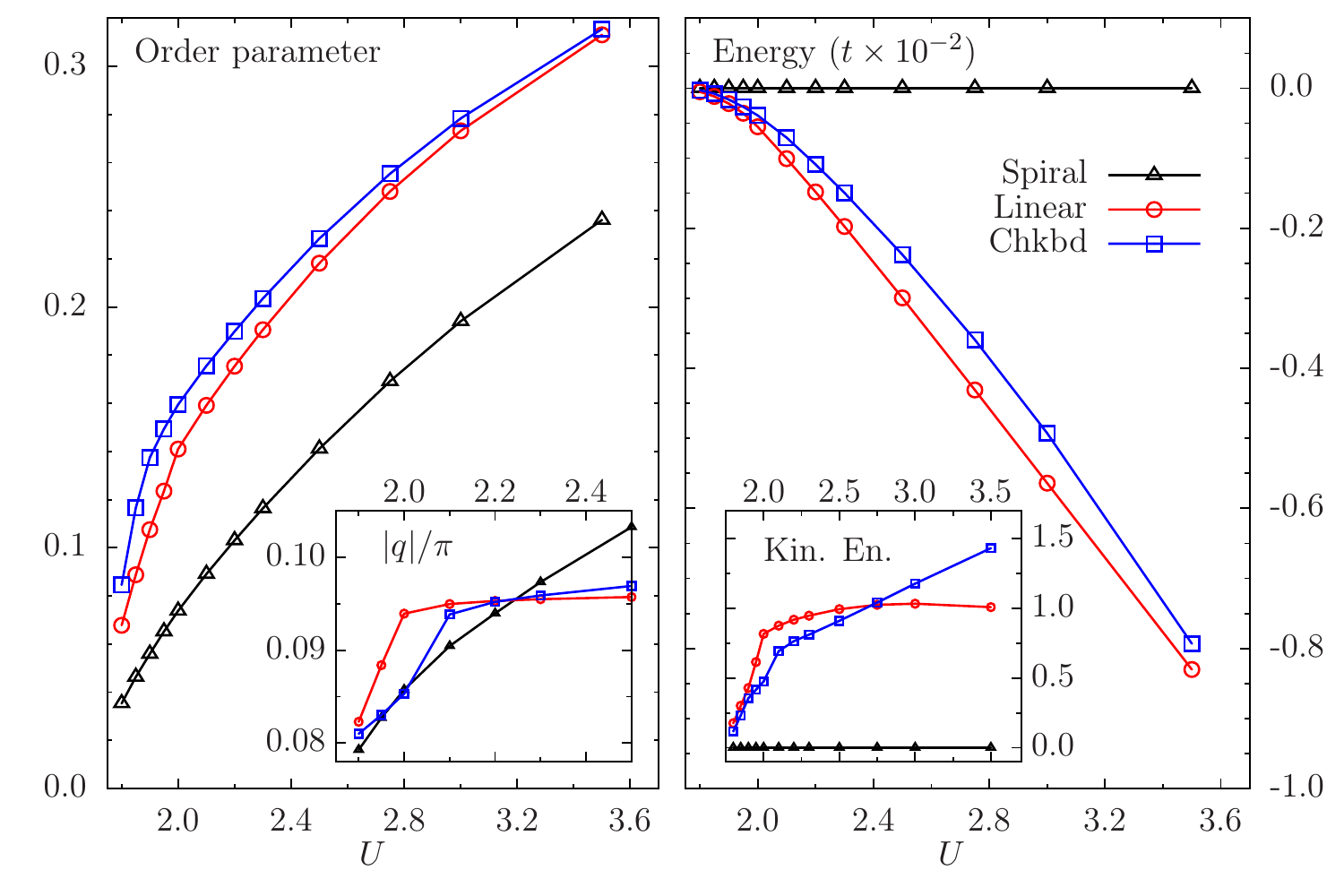}
\caption{Left Panel: local order parameter, $\max_i|\langle c_{i\uparrow} c_{i\downarrow} \rangle|$,
and leading wavevector (inset) versus $U$ for 
$\langle c_{i\uparrow} c_{i\downarrow} \rangle \propto e^{i q_x \cdot r_i}$ (Spiral), 
$\langle c_{i\uparrow} c_{i\downarrow} \rangle \propto \cos(q_x \cdot r_i)$ (Linear), 
$\langle c_{i\uparrow} c_{i\downarrow} \rangle \propto \cos(q_x \cdot r_i) +\cos(q_y r_i)$ (Chkbd) 
with $q_x=|q| (1,0)$ and $q_y=|q|(0,1)$. Right panel: relative energies of the
three phases. Data are for $n=0.95$ and $m/n=0.1$. Kinetic energies are in units of $t$.
}
\label{fig1}
\end{figure}

\begin{figure*}
\includegraphics[width=\textwidth]{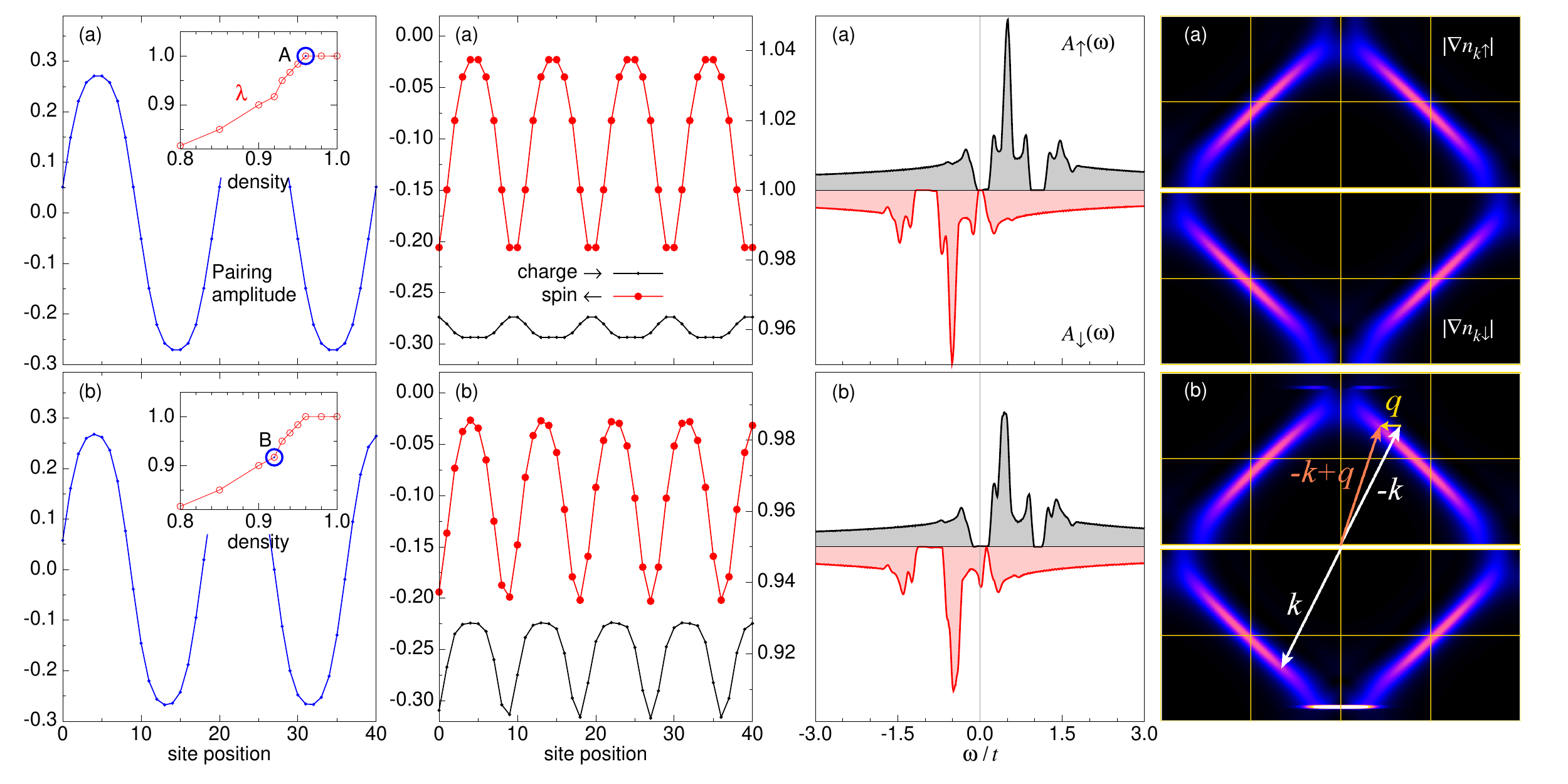}
\caption{
Local properties and gradient of the momentum distribution for 
$n_\uparrow - n_\downarrow=-0.05$ at two densities: (A) $n=0.96$ (top row) 
and (B) $n=0.93$ (bottom). In both cases, the finite-momentum of the pair, $q$, is in the $(1,0)$-direction.
The inset in the first column shows the 
evolution of $\lambda= m \pi / q$ as a function of density;
case A 
belongs to the commensurate regime with unit density 
of excess-spin particles per node, while case B is in the incommensurate regime.
The second column report the spin, $\langle n_{i\uparrow}-n_{i\downarrow}\rangle$,
and charge, $\langle n_{i\uparrow}+n_{i\downarrow}\rangle$,
density profiles in the direction of $q$.
The third column is the local density of states 
measured at site 0. In the fourth column, each panel shows two halves of the Fermi surface, 
for the $\uparrow$ (minority) and $\downarrow$ (majority) spins, respectively. The arrows in 
the figure indicate the pairing construction and applies to every pair across the FS, 
leading to one common 
$q$. 
}
\label{fig2}
\end{figure*}

\section{Symmetry of the order parameter}

\subsection{Determination}
\label{subsec:determ}

Given a set of symmetry-equivalent $q$ vectors and assuming a continuous phase transition,
linear response theory can be used to show that the onset 
of instabilities of the form
\begin{equation}
\Delta_i = \sum_q \Delta_q e^{i q \cdot r_i},
\end{equation}
must happen at exactly the same value of $U$, regardless of the choice
of $\Delta_q$. Below such value, mean field theory is guaranteed to return
a normal, spin-polarized Fermi liquid. In order to determine the correct form of order parameter,
we proceed as follows. We first determine $U_c$ and the associated
non-zero wave-vector $q_c$ using the single plane-wave form as this
allows for a quick exploration of phase space.
We find that $q_c$ is directed along any of 
the four, symmetry-equivalent  directions $(\pm1,0)$, $(0,\pm1)$
and, therefore, any linear combination of the four associated plane waves 
is a candidate ground state order parameter just above $U_c$. To resolve which one leads
to the largest lowering of energy, we proceed by solving the mean-field 
equations for the three cases corresponding to spiral ($\Delta_i\propto\exp(i q_x\cdot r_i)$), 
unidirectional ($\Delta_i\propto \cos(q_x\cdot r_i)$) and checkerboard ($\Delta_i\propto
\cos(q_x\cdot r_i)+\cos(q_y\cdot r_i)$) 
pairing density wave and track the evolution of $|q|$ as a function of $U$. 
Explicit calculations in large simulations cells  on the repulsive model 
(with simulated annealing starting with random initial fields\cite{Xu2011})
have shown that instabilities involving $q$-vectors in 
different, non-equivalent directions,  which the above approach would miss, are 
unlikely to occur in the range of $U$ considered here.

\subsection{Dependence on density and polarization}
Mean-field results\cite{Xu2011} on the repulsive model and
the particle-hole transformation relating the attractive and repulsive models 
imply the existence of the following properties at half-filling, i.e., when the average particle density is 
precisely one fermion per site:
1) a critical $U$ exists such that, above it,
the system develops a phase with an inhomogeneous order parameter 
2) the pairing order parameter is characterized by a wave vector $|q|=m\pi$
where $m=n_\uparrow - n_\downarrow$ is the average magnetization
3) both fermionic species have a gap in their single particle spectrum
4) as $U$ grows larger there is a transition from a pair-density-wave in the $(1,0)$
state to one in the $(1,1)$ direction 5) order can be arbitrarily broken into
a charge or a pairing instability or a combination of the two. This last 
point is a consequence of the possibility of breaking spin symmetry in any
of the three equivalent directions in the repulsive case. It implies that charge 
and pairing orders compete, in the sense that the larger one is, the 
smaller the other must be. 

Although there have been studies on several aspects
of the physics away from half-filling\cite{Wang2006}, the correct leading pairing 
wave vector (which requires a scan through supercell sizes) has not been determined. 
This has prevented a characterization of the nature of such 
phases. As a result the order of possible transitions and the related 
possibility of phase separation have not been resolved.
We address these questions here using the strategy outlined in the
previous subsection. A representative example is given in Figure \ref{fig1}
for the case with 
$m\equiv n_\uparrow - n_\downarrow = -0.095$ and $n=0.95$. 

To understand the effects due to the significantly different shape of the
Fermi surfaces in a lattice when compared to their circular counterparts in the continuum,
we repeat a similar analysis in different regimes 
of density and polarization, $n=0.95,\ 0.75$ and $0.5$ and $m/n=0.1$ and $0.4$, so as to have
a rather complete picture close and away from half-filling, at small and large polarization 
and in an interaction range that extends from $U_c$ up to $U=4$. 

The results  are summarized 
in Table \ref{tab1}.
They are consistent with earlier results addressing the physics of FFLO
phases in the continuum, which found checkerboard order close to $U_c$ thanks to
an expansion in powers of the order parameter\cite{Shimahara1997}, and show that for $n \ge 0.75$ 
lattice effects are strong enough to return unidirectional order independently of
polarization or proximity to $U_c$, as known to happen at and around $n=1.0$. 
Given the difficulty in observing unconventional
pairing in the continuum, the latter is the density regime where an
experimental realization of the FFLO state could be more feasible. We will
therefore focus on it in the following.

\begin{table}
\begin{center}
  \begin{tabularx}{\columnwidth}{ lXXX }
    \hline\hline
    $m/n$ & $n=0.95$ & $n=0.75$ & $n=0.5$ \\ \hline
    \multirow{3}{*}{$0.1\qquad$} & $U_c=1.8$     & $U_c=1.8$     & $U_c=2.0$ \\ 
                         & $q_c=0.071\pi$ & $q_c=0.078\pi$ & $q_c=0.066\pi$ \\ 
                         & Linear; Linear & Linear; Linear & Chkbd; Linear \\
    \multirow{3}{*}{$0.4\qquad$} & $U_c=3.6$     & $U_c=3.2$     & $U_c=2.8$ \\ 
                         & $q_c=0.34\pi$ & $q_c=0.30\pi$ & $q_c=0.27\pi$ \\ 
                         & Linear; Linear & Linear; Linear & Chkbd; Chkbd \\ \hline\hline
  \end{tabularx}
\end{center}
\caption{Critical parameter at the onset of order for a few densities ($n$) and
polarizations ($m/n$). The wavevector is directed along (1,0). The two entries in the last line in each table
cell describe, respectively, the type of order just above $U_c$ and at $U=4$ (see Fig.1's caption for detail).
}
\label{tab1}
\end{table}

\section{Character of the nodal phases} 

\subsection{Metal and band insulator at weak coupling}
Figure~\ref{fig2} characterizes the FFLO phase at $U=3t$ and for $m=0.05$ in two qualitatively 
different density regimes. In particular, for $n>0.95$, the wave vector is precisely 
determined by the magnetization via the same relation holding at half-filling,
$q=m\pi$. This commensurate regime is characterized by a density of {\em one} 
excess particle per node (of the order parameter) and consequent band-insulating behavior along the node. 
This is most clearly seen in the local density of states at the node, with both
species showing a gap at the Fermi energy. Correspondingly the gradient of the momentum
distribution shows no sharp lines indicative of the existence
of a Fermi surface. The commensurate regime here has, however, some important
distinctions from half-filling.
First, the interchangeability between pairing and charge orders is broken  
as soon as the average density deviates from $n=1$, and pairing emerges as the dominant order.
Second, the density is not perfectly uniform, as it is for the purely superfluid
phase at half-filling. The density is instead characterized by a weak modulation 
that reflects the different degrees of localization of the Andreev states for 
the two spin species. The density profile shows weak peaks at the nodes indicating that 
the majority-species nodal states have stronger localization. A similar phase was
found in the context of a two-dimensional array of tubes\cite{Parish2007}.

In the second density regime,  $n<0.95$, the majority 
spin species develop a finite density of states at the nodes of the order parameter.
Fermi arcs appear in $|\nabla n_{k\downarrow}|$ in the form of sharp lines.
The arc in Fig.~\ref{fig2}.B is of perfectly one-dimensional nature, indicating a complete
decoupling between the metallic states living at different nodes.
At larger polarization, the arcs
will more strongly resemble the underlying Fermi surface of the
non-interacting species. Finally, the density profile shows minima at the nodes, 
rather than the maxima observed in case A, as a direct consequence of the
gradual emptying of the majority-spin Andreev bands visible in the local density of states.
This is a potentially important experimental signature as it allows the characterization of the
nature of the nodal phase, metallic or insulating, via a static local property instead of 
a collective property such as the distance between nodes or the presence of Fermi arcs.

\begin{figure}
\includegraphics[width=\columnwidth]{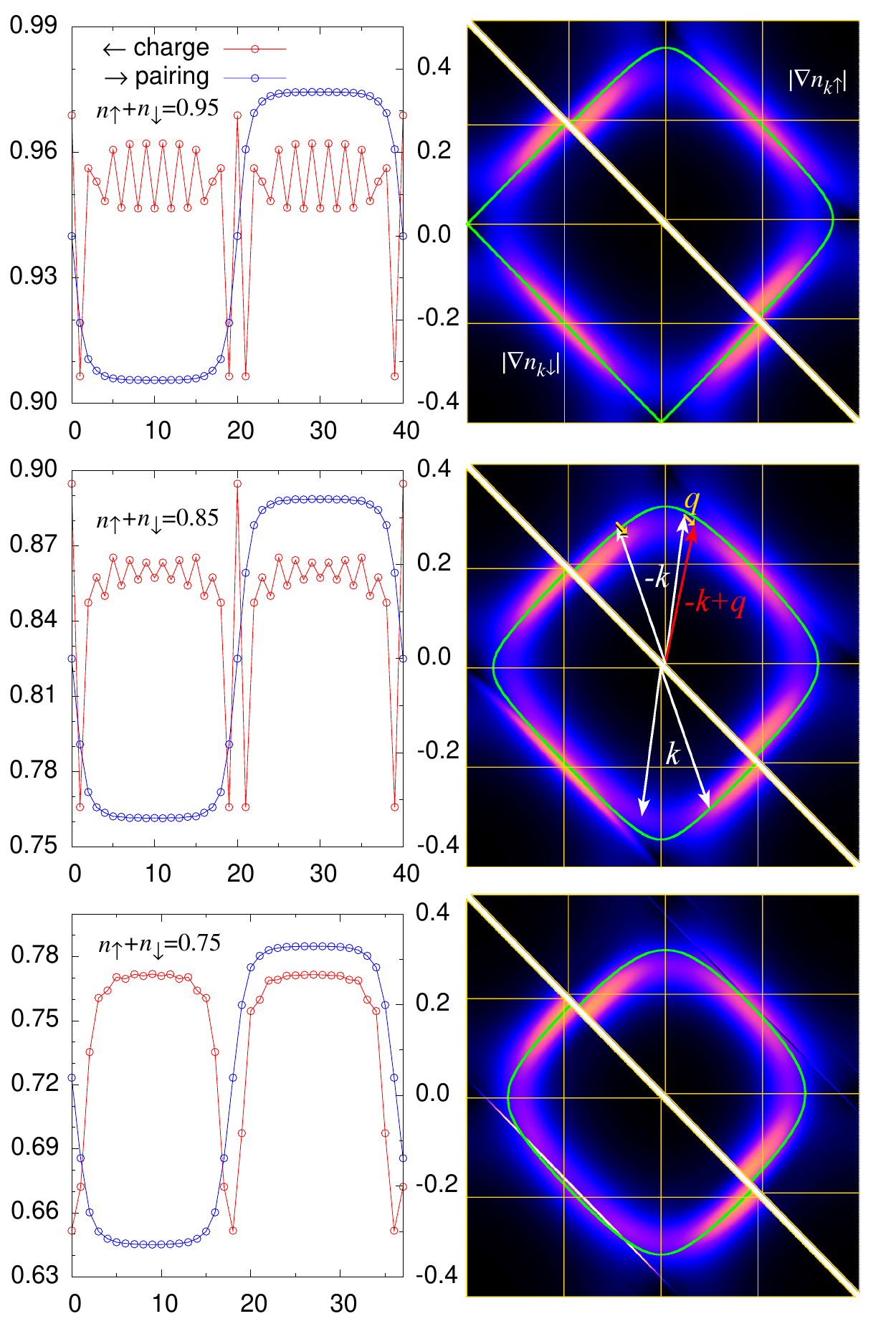}
\caption{Left panels: Evolution of charge $\langle n_{i\uparrow} + n_{i\downarrow}\rangle$ 
and pairing order as density is reduced, 
for $m=0.05$ along the $(1,0)$ direction. 
Right panels: 
corresponding evolution of the gradient of the
momentum distributions, and illustration of Fermi surface reconstruction and pairing. 
The pairing wave-vector, $q$, is in the $(-1,1)$-direction.
The solid (green) lines give the non-interacting Fermi surface.}
\label{fig3}
\end{figure}

\subsection{Supersolid phase at intermediate coupling}
Next we consider the phases at larger interaction strengths $U$.
Results on the repulsive model\cite{Xu2011} indicate that
the ordering wave vector switches to the $(1,1)$ direction at sufficiently
small $m$ and $U>3t$. 
We focus on $U=4t$ for an extensive and systematic 
examination of the properties of these phases.
Indeed unidirectional LO states along the diagonal direction are found; in addition, we find that the system
can accommodate both pairing and charge orders as {\em non-competing\/} instabilities away 
from half-filling. 
The real- and momentum-space properties of the phases are shown  in Fig.~\ref{fig3}.
In contrast to the (1,0)-direction states, 
pairing here is achieved by a more radical reconstruction of the
Fermi surface. The non-interacting surface of the majority spin is stretched, while that of 
the minority spin is compressed, along
the (-1,1) direction so that the reconstructed surfaces become 
``nested'' via $k\rightarrow -k+q$ as illustrated in the middle row of Fig.\ref{fig3}.
The  charge-density wave that develops at larger density is a consequence of the $(\pi,\pi)$ nesting 
along the $(1,1)$-direction which results from the reconstruction;
its amplitude increases with the amount of nested states. 
The charge order is complementary to pairing,  and further lowers the energy. 

The new phase discussed above has an important distinction from the supersolid phase 
that can occur at half-filling. There both charge and pairing orders are 
associated with wave-vectors coupling the same regions around the Fermi
surface: if pairing order happens at $q$, charge order must happen at $(\pi,\pi)-q$
and one can dial the form of the order parameter interpolating between
a purely superfluid and a purely charge-density wave state.
Away from half-filling, however,
pairing is characterized by a polarization dependent wave-vector $q$ 
while charge order appears at $(\pi,\pi)$ and
{\em cannot} be dialed away.

\begin{figure}
\includegraphics[width=\columnwidth]{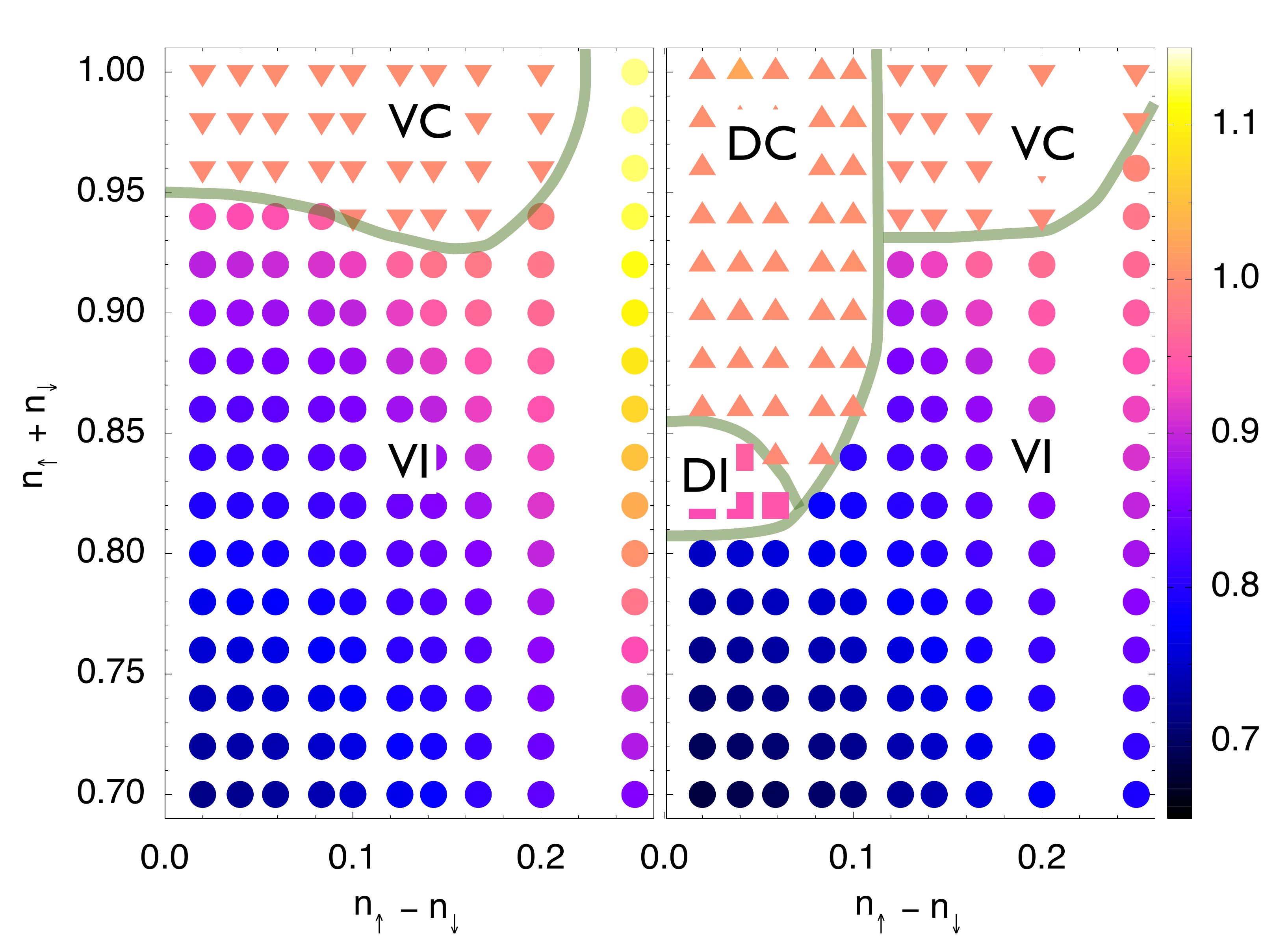}
\caption{$U/t=3$ (left) and $U/t=4$ (right) $T=0$ phase diagram. The color axis reports
values of $\lambda=m\pi/|q|$. Triangles represent the commensurate phases ($\lambda=1$). Up-triangles
have $q\propto (1,1)$ (DC=Diagonal commensurate) while down-triangles have $q\propto (1,0)$
(VC=Vertical Commensurate). Circles and square are Vertical and Diagonal Incommensurate
phases (VI and DI).}
\label{fig4}
\end{figure}

\section{Phase diagram}
Our results are summarized in the phase diagram of Fig.~\ref{fig4} for $U=3t$ and $U=4t$.
We found no evidence of diagonal supersolid order 
at $U=3t$ with a region of stability for the commensurate phase limited to small polarization
and density close to one. At $U=4t$, the diagonal wave vector 
is almost always commensurate with $m$, and pairing and charge order coexist in all but a 
small fraction of the phase diagram where the diagonal phase is the correct ground state.
This suggests that the charge-density wave play 
a small but important role in stabilizing diagonal order. 

\begin{figure}
\includegraphics[width=\columnwidth]{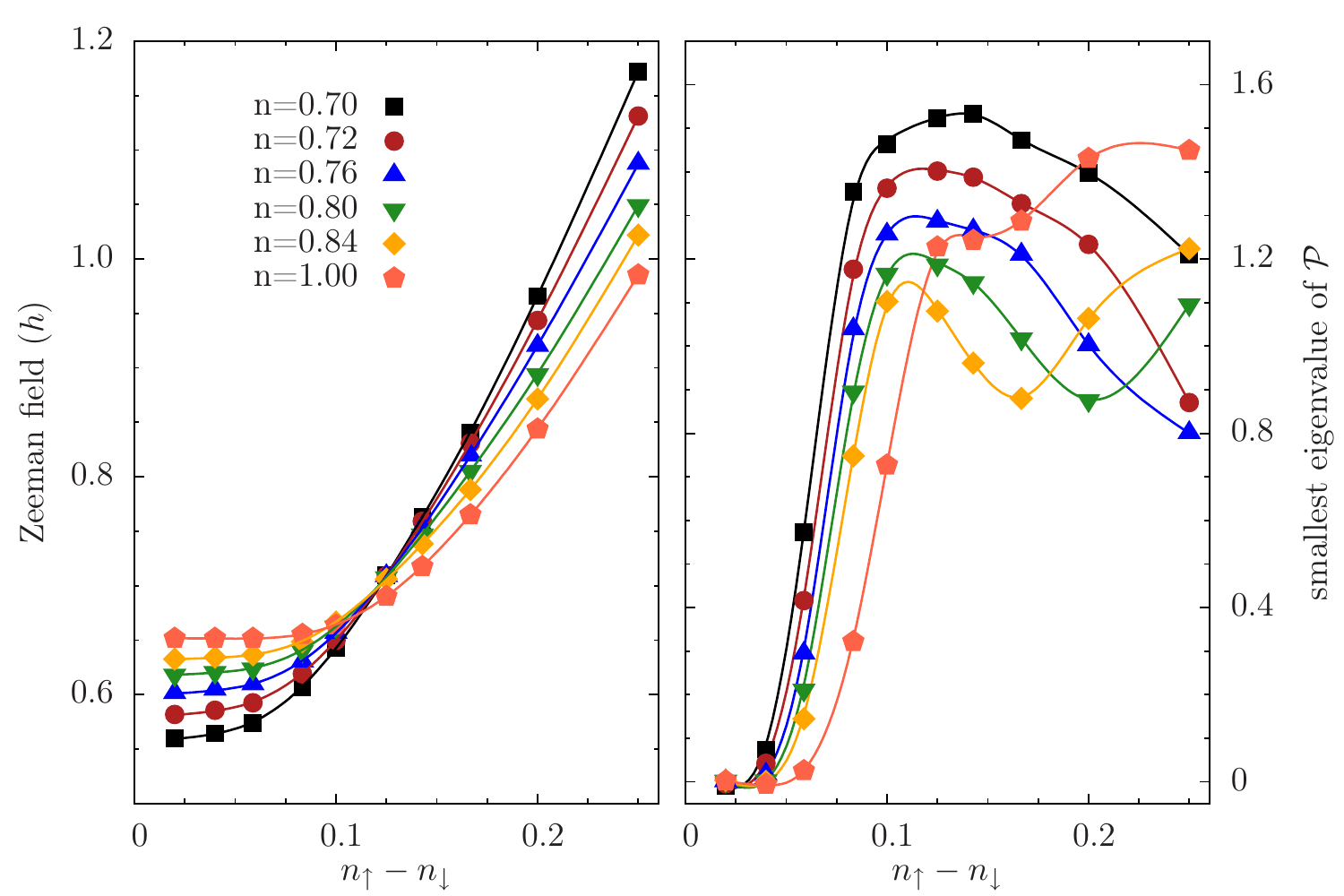}
\caption{Left: Zeeman field, $h$, as a function of magnetization, $m$, for different densities. 
At small magnetizations the Zeeman field reaches a plateau as explained in the main body of the
paper. Right: Smallest eigenvalue of the matrix $\mathcal{P}$. In the density/magnetization regime
considered this eigenvalue remains positive (although very small in correspondence of the Zeeman
field plateau) indicating no region of phase separation. Data are for $U=-3t$.}
\label{fig5}
\end{figure}

Because we have accurately determined the pair momenta and the correct symmetry of the order parameter 
as a function of magnetization and density, we can now
tackle the issue of phase separation in spin-imbalanced systems on a square lattice within
Hartree-Fock-Bogoliubov theory. 
To do this we check the convexity of the energy by diagonalizing 
\begin{equation}
\mathcal{P}=
\left[ 
\begin{array}{cc}
\frac{\partial \mu}{ \partial n} & \frac{\partial \mu}{ \partial m}\\
\frac{\partial h  }{ \partial n} & \frac{\partial   h}{ \partial m}\\
\end{array} 
\right]
\end{equation}
for the same set of densities and magnetizations reported in Fig.~\ref{fig4}. 
The values of $\mu$, $h$ and their derivatives are determined numerically from the Helmholtz 
free energy using the grid in Fig.\ref{fig4}. As a result, data in Fig.~\ref{fig5} have
numerical uncertainty that we estimated to be comparable to the symbol size. 
These data, summarizing the case of $U=3t$, do not suggest any tendency toward phase separation.
Although the smallest eigenvalue of $\mathcal{P}$ approaches $0$ at small magnetization
we do
not interpret this as a signal of incipient phase separation. Because the distance
between nodal lines is inversely proportional to the magnetization, 
the ground state at small magnetizations is characterized by the presence of essentially 
non-interacting domain walls where the excess spin particles reside. The ground state
energy is then simply determined by the energy to create one such wall multiplied by
the walls density. Because the latter is proportional to the magnetization,
the energy displays linear behavior in $m$ and causes $\partial h/\partial m$
to become vanishingly small. This, in turn, is responsible for the vanishing behavior
of one of the eigenvalues.
These results show that, contrary to previous conclusion based on restricted searches\cite{Igoshev2010}
on two-dimensional lattices and contrary to the widely accepted scenario in the dilute continuum, there 
is no phase separation in the true
Hartree-Fock-Bogoliubov ground state on a lattice in these parameter regimes.

Similar results hold for the phases at $U=4t$ when the different order parameter patterns are separately
considered. Obviously, in the global phase diagram, the different symmetry between diagonal and 
vertical phases implies that the transition is discontinuous and accompanied by phase separation.
We have not attempted an in-depth
study of how this affects the situation at $U=4t$. However, a simpler analysis based on 
considering phase separation in two phases having the same density and different magnetization 
leads to a narrow coexistence region ($\Delta m = 0.02$ at $n=1$) and makes it sensible to assume
that the broad feature of the phase diagram are, in fact, robust.

\section{Summary}
We have determined the type of phases that an imbalanced
population of two fermionic species with attractive interactions support in the two-dimensional square 
lattice. Their real- and momentum-space properties are quantitatively characterized.  
We find that unidirectional LO states are the 
most stable mean field solutions in a large range of parameter regimes, and
checkerboard order is only clearly favored at small density and large polarization.
At lower $U$, the finite-momentum pairing wave-vector is along $(1,0)$.
We have shown the insulating versus conducting nature of the nodal region of the superconducting order parameter, and its interplay with commensuration effects. 
Related to these effects, a new phase with supersolid order is seen when the ordering wave vector
is directed along the $(1,1)$ direction. 
Our results suggest that, besides time of flight or
Bragg spectroscopy,
the different local phases 
we discussed are also identifiable by their local density profiles. This, in turn, should help 
their real-space characterization in the presence of a confining potential.
Our results conclusively show that, within Hartree-Fock-Bogoliubov, there is no phase separation in the presence
of a lattice for the considered density and magnetization regimes.

\section*{Acknowledgments}

We acknowledge support from NSF (Grant no. ~DMR-1006217), DOE (DE-SC0008627), and ARO (Grant no.~56693-PH) and
computational support from the Center for Piezoelectrics by Design and by DOE leadership 
computing through an INCITE grant.

\bibliography{FFLO2D_references}

\begin{thebibliography}{14}
\expandafter\ifx\csname natexlab\endcsname\relax\def\natexlab#1{#1}\fi
\expandafter\ifx\csname bibnamefont\endcsname\relax
  \def\bibnamefont#1{#1}\fi
\expandafter\ifx\csname bibfnamefont\endcsname\relax
  \def\bibfnamefont#1{#1}\fi
\expandafter\ifx\csname citenamefont\endcsname\relax
  \def\citenamefont#1{#1}\fi
\expandafter\ifx\csname url\endcsname\relax
  \def\url#1{\texttt{#1}}\fi
\expandafter\ifx\csname urlprefix\endcsname\relax\def\urlprefix{URL }\fi
\providecommand{\bibinfo}[2]{#2}
\providecommand{\eprint}[2][]{\url{#2}}

\bibitem[{\citenamefont{Fulde and Ferrell}(1964)}]{Fulde1964}
\bibinfo{author}{\bibfnamefont{P.}~\bibnamefont{Fulde}} \bibnamefont{and}
  \bibinfo{author}{\bibfnamefont{R.~A.} \bibnamefont{Ferrell}},
  \bibinfo{journal}{Physical Review} \textbf{\bibinfo{volume}{135}},
  \bibinfo{pages}{A550} (\bibinfo{year}{1964}), ISSN \bibinfo{issn}{0031-899X},
  \urlprefix\url{http://link.aps.org/doi/10.1103/PhysRev.135.A550}.

\bibitem[{\citenamefont{Larkin and Ovchinnikov}(1965)}]{Larkin1965}
\bibinfo{author}{\bibfnamefont{A.}~\bibnamefont{Larkin}} \bibnamefont{and}
  \bibinfo{author}{\bibfnamefont{I.}~\bibnamefont{Ovchinnikov}},
  \bibinfo{journal}{Soviet Physics JETP} \textbf{\bibinfo{volume}{20}},
  \bibinfo{pages}{762 } (\bibinfo{year}{1965}),
  \urlprefix\url{http://www.citeulike.org/user/janpaniev/article/9341689}.

\bibitem[{\citenamefont{Sheehy and Radzihovsky}(2006)}]{Sheehy2006}
\bibinfo{author}{\bibfnamefont{D.~E.} \bibnamefont{Sheehy}} \bibnamefont{and}
  \bibinfo{author}{\bibfnamefont{L.}~\bibnamefont{Radzihovsky}},
  \bibinfo{journal}{Physical Review Letters} \textbf{\bibinfo{volume}{96}},
  \bibinfo{pages}{060401} (\bibinfo{year}{2006}),
  \urlprefix\url{http://arxiv.org/abs/cond-mat/0508430}.

\bibitem[{\citenamefont{Moreo and Scalapino}(2007)}]{Moreo2007}
\bibinfo{author}{\bibfnamefont{A.}~\bibnamefont{Moreo}} \bibnamefont{and}
  \bibinfo{author}{\bibfnamefont{D.}~\bibnamefont{Scalapino}},
  \bibinfo{journal}{Physical Review Letters} \textbf{\bibinfo{volume}{98}},
  \bibinfo{pages}{216402} (\bibinfo{year}{2007}), ISSN
  \bibinfo{issn}{0031-9007},
  \urlprefix\url{http://link.aps.org/doi/10.1103/PhysRevLett.98.216402}.

\bibitem[{\citenamefont{Wolak et~al.}(2012)\citenamefont{Wolak, Gr\'{e}maud,
  Scalettar, and Batrouni}}]{Wolak2012}
\bibinfo{author}{\bibfnamefont{M.~J.} \bibnamefont{Wolak}},
  \bibinfo{author}{\bibfnamefont{B.}~\bibnamefont{Gr\'{e}maud}},
  \bibinfo{author}{\bibfnamefont{R.~T.} \bibnamefont{Scalettar}},
  \bibnamefont{and} \bibinfo{author}{\bibfnamefont{G.~G.}
  \bibnamefont{Batrouni}}, \bibinfo{journal}{Physical Review A}
  p.~\bibinfo{pages}{15} (\bibinfo{year}{2012}), \eprint{1206.5050},
  \urlprefix\url{http://arxiv.org/abs/1206.5050}.

\bibitem[{\citenamefont{Chang and Zhang}(2010)}]{Chang2010}
\bibinfo{author}{\bibfnamefont{C.-C.} \bibnamefont{Chang}} \bibnamefont{and}
  \bibinfo{author}{\bibfnamefont{S.}~\bibnamefont{Zhang}},
  \bibinfo{journal}{Phys. Rev. Lett.} \textbf{\bibinfo{volume}{104}},
  \bibinfo{pages}{116402} (\bibinfo{year}{2010}),
  \urlprefix\url{http://link.aps.org/doi/10.1103/PhysRevLett.104.116402}.

\bibitem[{\citenamefont{Koponen et~al.}(2006)\citenamefont{Koponen, Kinnunen,
  Martikainen, Jensen, and T\"{o}rm\"{a}}}]{Koponen2006}
\bibinfo{author}{\bibfnamefont{T.}~\bibnamefont{Koponen}},
  \bibinfo{author}{\bibfnamefont{J.}~\bibnamefont{Kinnunen}},
  \bibinfo{author}{\bibfnamefont{J.-P.} \bibnamefont{Martikainen}},
  \bibinfo{author}{\bibfnamefont{L.~M.} \bibnamefont{Jensen}},
  \bibnamefont{and}
  \bibinfo{author}{\bibfnamefont{P.}~\bibnamefont{T\"{o}rm\"{a}}},
  \bibinfo{journal}{New Journal of Physics} \textbf{\bibinfo{volume}{8}},
  \bibinfo{pages}{179} (\bibinfo{year}{2006}), ISSN \bibinfo{issn}{1367-2630},
  \urlprefix\url{http://stacks.iop.org/1367-2630/8/i=9/a=179?key=crossref.58b7%
381cef82e44d246df7a855d2b7dd}.

\bibitem[{\citenamefont{Koponen et~al.}(2007)\citenamefont{Koponen, Paananen,
  Martikainen, and T\"{o}rm\"{a}}}]{Koponen2007}
\bibinfo{author}{\bibfnamefont{T.~K.} \bibnamefont{Koponen}},
  \bibinfo{author}{\bibfnamefont{T.}~\bibnamefont{Paananen}},
  \bibinfo{author}{\bibfnamefont{J.-P.} \bibnamefont{Martikainen}},
  \bibnamefont{and}
  \bibinfo{author}{\bibfnamefont{P.}~\bibnamefont{T\"{o}rm\"{a}}},
  \bibinfo{journal}{Physical Review Letters} \textbf{\bibinfo{volume}{99}},
  \bibinfo{pages}{120403} (\bibinfo{year}{2007}),
  \urlprefix\url{http://link.aps.org/doi/10.1103/PhysRevLett.99.120403}.

\bibitem[{\citenamefont{Loh and Trivedi}(2009)}]{Loh2009}
\bibinfo{author}{\bibfnamefont{Y.~L.} \bibnamefont{Loh}} \bibnamefont{and}
  \bibinfo{author}{\bibfnamefont{N.}~\bibnamefont{Trivedi}},
  \bibinfo{journal}{arXiv09070679v1 condmatquantgas}
  \textbf{\bibinfo{volume}{43210}}, \bibinfo{pages}{4} (\bibinfo{year}{2009}),
  \urlprefix\url{http://arxiv.org/abs/0907.0679}.

\bibitem[{\citenamefont{Wang et~al.}(2006)\citenamefont{Wang, Chen, Hu, and
  Ting}}]{Wang2006}
\bibinfo{author}{\bibfnamefont{Q.}~\bibnamefont{Wang}},
  \bibinfo{author}{\bibfnamefont{H.-Y.} \bibnamefont{Chen}},
  \bibinfo{author}{\bibfnamefont{C.-R.} \bibnamefont{Hu}}, \bibnamefont{and}
  \bibinfo{author}{\bibfnamefont{C.~S.} \bibnamefont{Ting}},
  \bibinfo{journal}{Phys. Rev. Lett.} \textbf{\bibinfo{volume}{96}},
  \bibinfo{pages}{117006} (\bibinfo{year}{2006}),
  \urlprefix\url{http://link.aps.org/doi/10.1103/PhysRevLett.96.117006}.

\bibitem[{\citenamefont{Xu et~al.}(2011)\citenamefont{Xu, Chang, Walter, and
  Zhang}}]{Xu2011}
\bibinfo{author}{\bibfnamefont{J.}~\bibnamefont{Xu}},
  \bibinfo{author}{\bibfnamefont{C.-C.} \bibnamefont{Chang}},
  \bibinfo{author}{\bibfnamefont{E.~J.} \bibnamefont{Walter}},
  \bibnamefont{and} \bibinfo{author}{\bibfnamefont{S.}~\bibnamefont{Zhang}},
  \bibinfo{journal}{Journal of physics. Condensed matter : an Institute of
  Physics journal} \textbf{\bibinfo{volume}{23}}, \bibinfo{pages}{505601}
  (\bibinfo{year}{2011}), ISSN \bibinfo{issn}{1361-648X},
  \urlprefix\url{http://www.ncbi.nlm.nih.gov/pubmed/22127010}.

\bibitem[{\citenamefont{Shimahara}(1998)}]{Shimahara1997}
\bibinfo{author}{\bibfnamefont{H.}~\bibnamefont{Shimahara}},
  \bibinfo{journal}{Journal of the Physical Society of Japan}
  \textbf{\bibinfo{volume}{67}}, \bibinfo{pages}{736} (\bibinfo{year}{1998}),
  \urlprefix\url{http://arxiv.org/abs/cond-mat/9711017}.

\bibitem[{\citenamefont{Parish et~al.}(2007)\citenamefont{Parish, Baur,
  Mueller, and Huse}}]{Parish2007}
\bibinfo{author}{\bibfnamefont{M.~M.} \bibnamefont{Parish}},
  \bibinfo{author}{\bibfnamefont{S.~K.} \bibnamefont{Baur}},
  \bibinfo{author}{\bibfnamefont{E.~J.} \bibnamefont{Mueller}},
  \bibnamefont{and} \bibinfo{author}{\bibfnamefont{D.~A.} \bibnamefont{Huse}},
  \bibinfo{journal}{Phys. Rev. Lett.} \textbf{\bibinfo{volume}{99}},
  \bibinfo{pages}{250403} (\bibinfo{year}{2007}),
  \urlprefix\url{http://link.aps.org/doi/10.1103/PhysRevLett.99.250403}.

\bibitem[{\citenamefont{Igoshev et~al.}(2010)\citenamefont{Igoshev, Timirgazin,
  Katanin, Arzhnikov, and Irkhin}}]{Igoshev2010}
\bibinfo{author}{\bibfnamefont{P.~A.} \bibnamefont{Igoshev}},
  \bibinfo{author}{\bibfnamefont{M.~A.} \bibnamefont{Timirgazin}},
  \bibinfo{author}{\bibfnamefont{A.~A.} \bibnamefont{Katanin}},
  \bibinfo{author}{\bibfnamefont{A.~K.} \bibnamefont{Arzhnikov}},
  \bibnamefont{and} \bibinfo{author}{\bibfnamefont{V.~Y.}
  \bibnamefont{Irkhin}}, \bibinfo{journal}{Phys. Rev. B}
  \textbf{\bibinfo{volume}{81}}, \bibinfo{pages}{094407}
  (\bibinfo{year}{2010}),
  \urlprefix\url{http://link.aps.org/doi/10.1103/PhysRevB.81.094407}.

\end{thebibliography}

\end{document}